\documentclass[aps,superscriptaddress,showpacs,amsmath,amssymb,longbibliography,twocolumn,floatfix]{revtex4-1}
\usepackage{times}
\usepackage{setspace}
\usepackage[margin=0.7in]{geometry}
\usepackage{comment}
\usepackage[utf8]{inputenc}
\usepackage{graphicx}
\usepackage{color}
\usepackage{xcolor}
\usepackage{soul}
\usepackage{environ} 
\usepackage{float} 
\usepackage{mathtools}
\usepackage{natbib}
\usepackage[pdftex, colorlinks=true, linkcolor=blue, citecolor=blue,urlcolor=blue]{hyperref}

\newcommand{\selfenergy}{[\Sigma_{n\mathbf{k}}]}

\begin{document}

\title{Thermalization of photoexcited carriers in two-dimensional transition metal dichalcogenides \\ and internal quantum efficiency of van der Waals heterostructures}

\author{Dinesh Yadav}
\affiliation{Okinawa Institute of Science and Technology Graduate University, Onna-son, Okinawa 904-0495, Japan}
\affiliation{Department of Physics, University of Konstanz, D-78457 Konstanz,
  Germany}
\author{Maxim Trushin}
\affiliation{Centre for Advanced 2D Materials, National University of Singapore, 6 Science Drive 2, Singapore 117546}
\author{Fabian Pauly}
\affiliation{Okinawa Institute of Science and Technology Graduate University, Onna-son, Okinawa 904-0495, Japan}
\affiliation{Department of Physics, University of Konstanz, D-78457 Konstanz, Germany}
  
\date{\today} 
  
\begin{abstract}
Van der Waals semiconductor heterostructures could be a platform to harness hot photoexcited carriers in the next generation of optoelectronic and photovoltaic devices. The internal quantum efficiency of hot-carrier devices is determined by the relation between photocarrier extraction and thermalization rates. Using \textit{ab-initio} methods we show that the photocarrier thermalization time in single-layer transition metal dichalcogenides strongly depends on the peculiarities of the phonon spectrum and the electronic spin-orbit coupling. In detail, the lifted spin degeneracy in the valence band suppresses the hole scattering on acoustic phonons, slowing down the thermalization of holes by one order of magnitude as compared to electrons. Moreover, the hole thermalization time behaves differently in MoS$_2$ and WSe$_2$ because spin-orbit interactions differ in these seemingly similar materials. We predict that the internal quantum efficiency of a tunneling van der Waals semiconductor heterostructure depends qualitatively on whether MoS$_2$ or WSe$_2$ is used.
\end{abstract}

 \maketitle

\section {Introduction}
Utilizing high-energy carriers in photovoltaic devices could improve light-to-energy conversion efficiencies \cite{nelson2003physics}. Despite recent progress with hot-carrier generation and injection in plasmonic nanostructures \cite{clavero2014plasmon}, the conventional semiconductor solar cells still demonstrate a superior efficiency without need for nanoscale fabrication. Alternative approaches to the problem either facilitate hot-electron transfer, {\it e.g.} from a chemically modified surface of lead selenide to titanium oxide \cite{Science2010Ti02}, or extend photocarrier lifetimes, {\it e.g.} in some perovskites \cite{perovskite-abinitio,perovskite-exper}. Van der Waals (vdW) semiconductor heterostructures formed from atomically thin two-dimensional (2D) crystals \cite{NovoselovReview2016} might represent a suitable platform to take advantage of both phenomena, {\it i.e.} ultrafast photocarrier extraction and slow photocarrier thermalization. 
    
\begin{figure}[!tb] \centering{}\includegraphics[width=1.0\columnwidth]{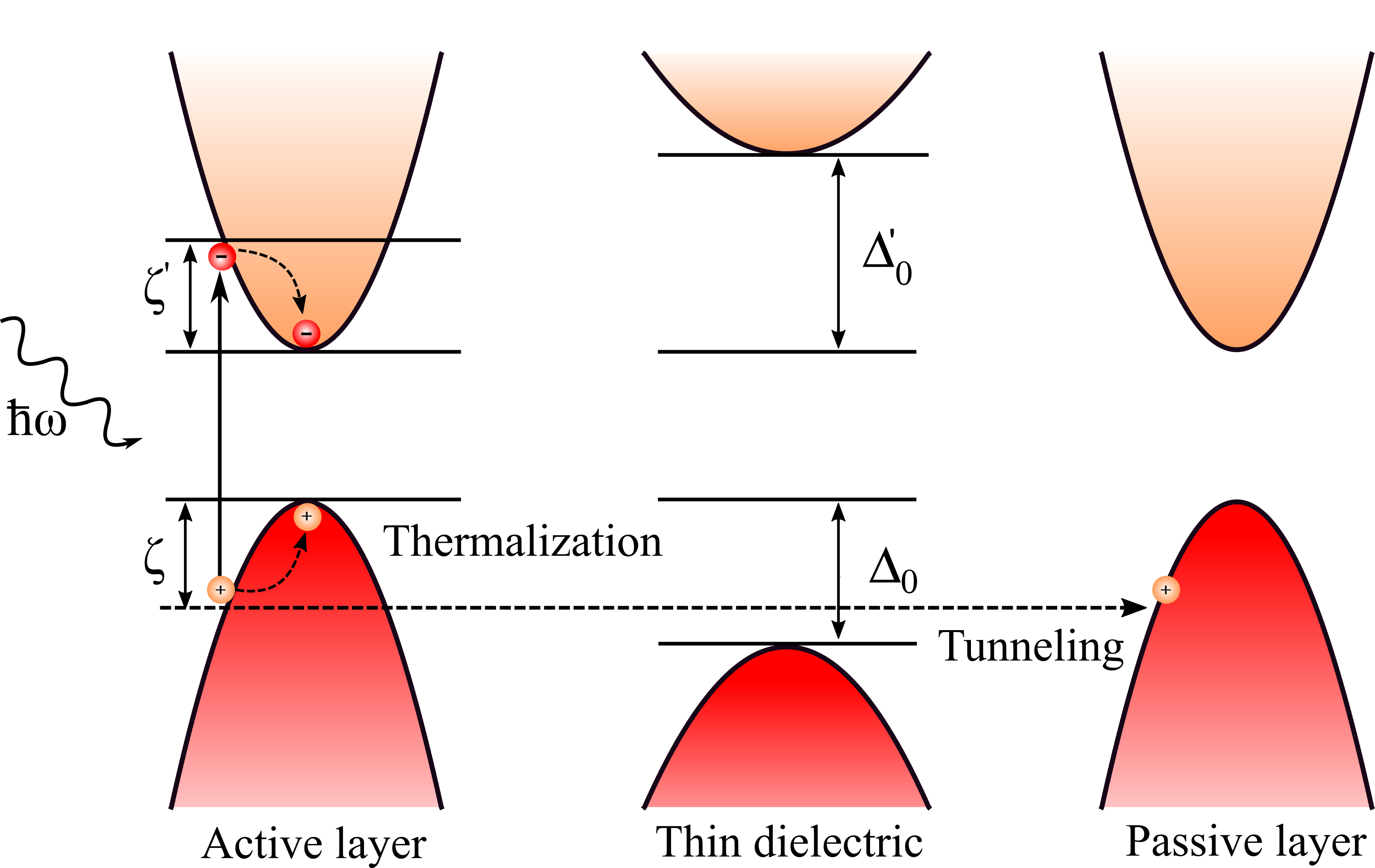}
\caption{Photocarrier excitation, thermalization, and tunneling in a
semiconductor--dielectric--semiconductor vdW heterostructure.
A weak bias (not shown) is applied to create a current across the layers.
The internal quantum efficiency is determined by the ratio between thermalization and tunneling rates. Here, we consider transport through the valence bands because of the longer thermalization time for photoexcited holes.} \label{fig:heterostructure}
\end{figure}

The possibility to assemble vdW heterostructures layer by layer allows to tune electron transfer across interfaces in the out-of-plane dimension. This interlayer charge transfer directly competes with intralayer photocarrier thermalization. The scheme has first been realized in a graphene--boron-nitride--graphene vdW heterostructure \cite{ma2016tuning}, where the interlayer tunneling time ranges from 1~fs to 1~ps depending on bias voltage. Since the photocarrier thermalization time in graphene spans the interval between 10~fs and 10~ps depending on carrier concentration and excitation energy \cite{NL2014epi,PRB2015fast,PRL2016slow,PRB2015doping,PRB2016collinear,graphene_bottleneck}, there is a parameter range within which high-energy photocarrier extraction is feasible. Substituting graphene by a 2D transition metal dichalcogenide (TMDC) in a stack \cite{Review_Butler2013,Review_beyond_graphene,Review_electronic_optoelectronic_TMDCs,Application_mos2} opens new perspectives thanks to the possibility to create a diode configuration \cite{2Dphotovoltaic} and to realize stronger light-matter interactions \cite{2DTMDCgraphene2013}. There are already several reports on the fabrication of TMDC-based vdW heterostructures and their optoelectronic properties \cite{vdW2013not2D,vdW2017vu,vdW2017eda}. The high-energy photocarriers may be filtered by means of a boron-nitride layer, constituting a barrier for thermalized electrons and holes with low energy \cite{vdW2017vu,vdW2017eda}. Since the photocarrier thermalization time is sensititive to the electron and phonon spectra of each semiconductor, it is important to compare this thermalization time with the interlayer tunneling time: The high-energy photocarrier transport may not be feasible, if the former is too short. The relevant processes are sketched in Fig.~\ref{fig:heterostructure}. Note that we do not distinguish between the terms "hot" or "high-energy" carriers, but use them interchangably.

Light-matter interaction in 2D TMDCs has recently been explored in time-resolved photoluminescence and transient absorption spectroscopy experiments using either linearly or circularly polarized light targeting exciton and carrier dynamics \cite{Ultrafast_carrier_thermalization_MoS2_1,Mos2_valleypolaization_1,Exciton_dynamics_mos2,EEA_mos2,EEA_tmd,Exciton_dynamics_wse2,Carrier_polarisation_dynamics_mos2, Defect_recombination_mos2, Valley_relaxation_wse2, Valley_depolarization_wse2, Valleytronics_wse2, Ultrafast_carrier_thermalization_MoS2_2,Ultrafast_carrierdynamics_wse2, Carrier_dynamics_free_carriers, Linewidths_exciton_TMDS, Electronic_phononic_ws2}. Experimental and theoretical studies of 2D TMDCs have largely focused on bound states for excitation energies below the electronic band gap, reporting exciton lifetimes \cite{Bernardi_excitons, Exciton_dynamics_wse2,Carrier_polarisation_dynamics_mos2}, excitonic linewidths and diffusion rates \cite{Linewidths_exciton_TMDS, Electronic_phononic_ws2}, exciton-exciton annihilation \cite{EEA_mos2,EEA_tmd}, and exciton surface defect trapping \cite{Exciton_dynamics_mos2}. Valley-resolved carrier dynamics has also been investigated around the band edge, and long valley depolarisation times starting from a few ps for MoS$_2$ \cite{Carrier_polarisation_dynamics_mos2,Mos2_valleypolaization_1} to a few ns for WSe$_2$ \cite{Valley_relaxation_wse2,Valley_depolarization_wse2,Valleytronics_wse2} have been measured. Free photoexcited carrier thermalization in 2D TMDCs, arising for excitations above the band gap, has been studied much less comprehensively \cite{Ultrafast_carrier_thermalization_MoS2_1, Ultrafast_carrier_thermalization_MoS2_2,Ultrafast_carrierdynamics_wse2, Carrier_dynamics_free_carriers,Momentum_resolvedEP_wse2, Defect_recombination_mos2}. \textit{Ab-initio} studies have shown that the carrier-carrier scattering is efficient far away from the band edges due to a quadratic increase of the available phase space \cite{Neglect_e_e_interaction_TMDC, Bernardi_silicon}. For a clean sample with no defects, the interaction with phonons thus represents the most important nonradiative channel for thermalization and cooling of photoexcited carriers in the vicintiy of the valence and conduction band edges \cite{Exciton_dynamics_mos2, EP_interaction,Momentum_resolvedEP_wse2} and requires a deeper understanding.

In what follows, we use an approach already applied successfully to bulk and 2D materials \cite{Bernardi_silicon,Bernardi_GaAs,graphene_bottleneck,CdTe}. For instance we have demonstrated in our recent work that phonon emission by photocarriers in graphene is strongly suppressed at low excess energies of about 100~meV \cite{graphene_bottleneck}. This phenomenon occurs because of high optical phonon frequencies due to strong carbon-carbon bonding and has been observed experimentally as a thermalization bottleneck \cite{PRL2016slow}. In the present work we study the influence of spin-orbit coupling (SOC) on photocarrier thermalization in single-layer TMDCs and additionally relate intralayer thermalization rates to interlayer tunneling rates in vdW heterostructures, important for future optoelectronic devices with improved internal quantum efficiency (IQE). For this purpose we make use of density functional theory (DFT) and density functional perturbation theory (DFPT) to determine the electronic and phononic spectra of two 2D TMDC representatives, namely MoS$_2$ and WSe$_2$. Furthermore, we compute band- and momentum-dependent scattering rates $\tau_{n\mathbf{k}}^{-1}$ for a given electronic band $n$ and wave vector $\mathbf{k}$. Applying the Boltzmann equation in the relaxation time approximation (RTA) we extract the total thermalization time $\tau_\mathrm{th}$ as a function of excess energy and temperature. In our parameter-free \textit{ab-initio} modeling, we take relativistic SOC and all the optical and acoustical phonon branches in the first Brillouin zone (BZ) into account to provide a reliable description of electron-phonon scattering events. We show that photocarrier thermalization is slowed down by SOC-induced band splitting near valence band maxima and conduction band minima. Similar to interlayer transport, the thermalization occurs faster for photocarriers excited farther away from the band edges. We find, however, that the hole thermalization rate increases with excess energy in MoS$_2$ much more rapidly than in WSe$_2$. We therefore expect that the IQE of the tunneling device shown in Fig.~\ref{fig:heterostructure} will strongly depend on whether MoS$_2$ or WSe$_2$ is employed.

 \section{Theoretical methods}
    
\subsection{\emph{Ab-initio} theory for electronic and phononic properties}\label{subsec:method-ab-initio}

We determine electronic and phononic properties of MoS$_2$ and WSe$_2$ monolayers using DFT within the local density approximation (LDA) as implemented in \textsc{Quantum Espresso} \citep{QUANTUMESPRESSO}. Six outermost electrons of each transition metal (Mo, W) and chalcogen (S, Se) are treated explicitly as valence electrons, while the remaining core electrons are included through norm-conserving Troullier-Martins pseudopotentials with relativistic corrections \cite{TM_pseudopotential}. We employ a plane-wave basis set with a kinetic energy cutoff of 70~Ry and a charge density cutoff of 280~Ry. Unit cells of MoS$_2$ and WSe$_2$ monolayers are optimized with the help of the Broyden-Fletcher-Goldfarb-Shanno algorithm, neglecting SOC until the net force on atoms is less than $10^{-6}$~Ry/a.u.\ and total energy changes are below $10^{-8}$~Ry. The monolayers are placed in a cell with a vacuum that separates periodic images by 18~\textup{\AA} to avoid artificial interactions in the out-of-plane direction. After geometry opimization we calculate the ground state density with and without SOC, sampling the BZ with a $45\times45\times1$ $\Gamma$-centered $\mathbf{k}$ grid. We evaluate the phonon dispersion of MoS$_2$ and WSe$_2$ monolayers through DFPT \citep{DFPT_RevModPhys.73.515} with and without SOC, employing a $12\times12\times1$ $\mathbf{q}$ grid to obtain phonon dynamical matrices. Next we construct localized Wannier functions from plane-wave eigenfunctions. By using the Wannier function interpolation scheme, we obtain electronic eigenenergies, dynamical matrices and electron-phonon matrix elements on desired grids in the BZ \citep{Wannierorbitals, EP_wannier, EPW_package_PONCE2016116}. 

\begin{figure}[!tb] \centering{}\includegraphics[width=1.0\columnwidth]{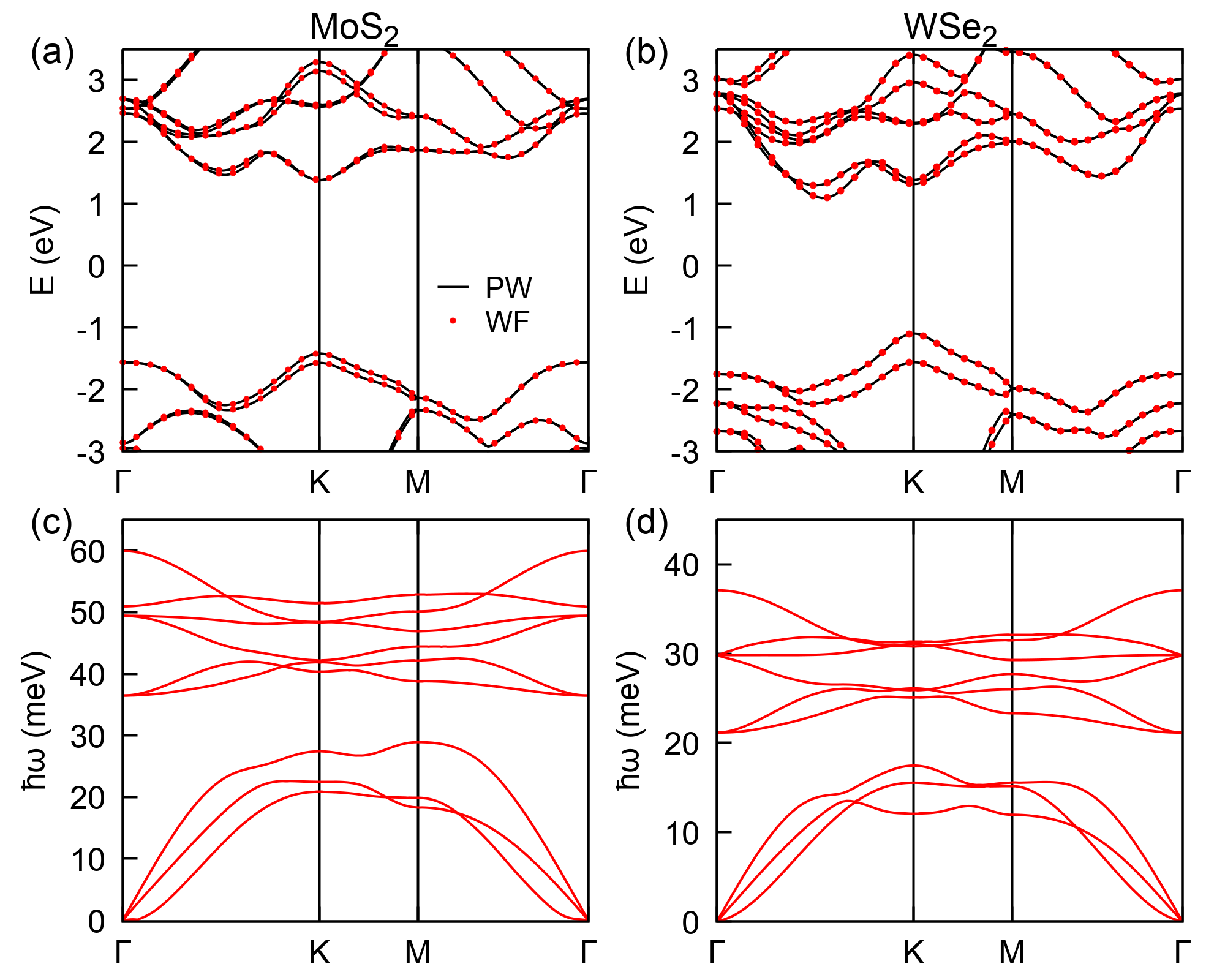}
\caption{Electronic band structure for monolayer (a) MoS$_2$ and (b) WSe$_2$, calculated with plane wave (PW) and Wannier functions (WF) basis sets. Phonon band structure for monolayer (c) MoS$_2$ and (d) WSe$_2$. Both electronic and phononic band structures take the relativistic SOC into account.} 
\label{fig:bandstruct-e-ph}
\end{figure}

After structural relaxation the in-plane lattice constants $|\textbf{a}|$ for MoS$_2$ and WSe$_2$ monolayers turn out to be 3.16~\textup{\AA} and 3.27~\textup{\AA}, respectively, which is in good agreement with  experiment and previous \textit{ab-initio} calculations \cite{latticeconstant_experiment,latticeconstatnt_bandgap_theory}. The electronic band structure is displayed in Fig.~\ref{fig:bandstruct-e-ph}(a,b). We find a direct quasiparticle band gap of 1.74~eV at the K point in the BZ for MoS$_2$, while WSe$_2$ exhibits an indirect gap of 1.44~eV with the valence band maximum at the K point and the conduction band minimum located on the $\Gamma$-K direction. As DFT is known to underestimate band gaps, we apply a rigid shift to the unoccoupied conduction bands to arrive at values of 2.82 and 2.42~eV for MoS$_2$ and WSe$_2$, respectively, which are consistent with the GW approximation \cite{Excitons_MX2_BSE, Exciton_louie}. This shift modifies the band gap values but does not affect the time evolution of the hot carriers to be discussed later. This assumption is valid because the electronic band gap is much larger than the highest phonon energy, and electrons and holes thus thermalize independently. Fig.~\ref{fig:bandstruct-e-ph}(a,b) furthermore shows that the Wannier-function-interpolated band structure matches exactly with the band structure determined with the plane wave basis, confirming the high quality of the localized basis set. Due to three atoms in the unit cell MoS$_2$ and WSe$_2$ feature three acoustical and six optical modes of vibration, as it is visible from the phononic band structure in Fig.~\ref{fig:bandstruct-e-ph}(c,d). The highest frequencies of 60 and 37~meV for MoS$_2$ and WSe$_2$ monolayers occur at the $\Gamma$ point, respectively, and these values agree with previously reported ones \cite{phonon_MX2,Phonon_sl_MoS2}. An energy gap separates acoustical and optical branches, which will be exploited in the analysis of thermalization times further below.
    
Having determined electronic and phononic band structures, we evaluate the electronic self-energy $\Sigma_{n\mathbf{k}}(T)$ in the lowest order of the electron-phonon interaction for band $n$ and wave vector $\mathbf{k}$ at temperature $T$ using the \textsc{EPW} code \citep{EPW_package_PONCE2016116}. It is expressed as 
\begin{widetext}
    \begin{equation}
        \Sigma_{n\mathbf{k}}(T) = \sum_{m,p}\int_{\text{BZ}}\frac{d^3q}{\Omega_{\text{BZ}}}\vert
        g_{mn,p}(\mathbf{k},\mathbf{q})\vert^{2}\Bigg[\frac{N_{p\mathbf{q}}(T)+f^{(0)}_{m\mathbf{k+q}}(T)}{\varepsilon_{n\mathbf{k}}-(\varepsilon_{m\mathbf{k+q}}-\varepsilon_{\text{F}})+\hbar\omega_{p\mathbf{q}}+\text{i}\eta}\\
        +\frac{N_{p\mathbf{q}}(T)+1-f^{(0)}_{m\mathbf{k+q}}(T)}{\varepsilon_{n\mathbf{k}}-(\varepsilon_{m\mathbf{k+q}}-\varepsilon_{\text{F}})-\hbar\omega_{p\mathbf{q}}+\text{i}\eta}\Bigg].\label{eq:self-energy}
    \end{equation}
\end{widetext}
Here, $\hbar\omega_{p\mathbf{q}}$ is the energy of the phonon of branch $p$ at wave vector $\mathbf{q}$, $\varepsilon_{\text{F}}=0$ is the Fermi energy, $f^{(0)}_{n\mathbf{k}}(T)=1/[\exp(\frac{\varepsilon_{n\textbf{k}}-\varepsilon_{\text{F}}}{k_{\text{B}}T})+1]$ is the Fermi-Dirac distribution, $N_{\hbar\omega_{p\mathbf{q}}}(T)=1/[\exp(\frac{\hbar\omega_{p\mathbf{q}}}{k_{\text{B}}T})-1]$ is the Bose function, $\Omega_{\text{BZ}}$ is the volume of the BZ, and $\eta=20~\text{meV}$ is a small broadening parameter. The electron-phonon matrix elements are defined as \citep{EPW_package_PONCE2016116}
\begin{equation}
  g_{mn,p}(\mathbf{k,q})=\frac{1}{\sqrt{2\omega_{p\mathbf{q}}}}\left<m\mathbf{k+q}|\partial_{p\mathbf{q}}V|n\mathbf{k}\right>,
\end{equation}
describing transitions between the Kohn-Sham states $|n\textbf{k}\rangle$ and $|m\textbf{k}+\mathbf{q}\rangle$ mediated by the phonon $p\mathbf{q}$. Here, $\partial_{p\mathbf{q}}V$ is the derivative of the self-consistent Kohn-Sham potential with respect to displacements of nuclei along the phonon mode $p\mathbf{q}$. The first term in the brackets of Eq.~(\ref{eq:self-energy}) arises from the absorption of phonons and the second one from their emission. To compute $\Sigma_{n\mathbf{k}}(T)$, we interpolate electronic and vibrational states on fine $300\times300\times1$ \textbf{k} and \textbf{q} grids, which we find sufficient to accurately map the first BZ and to converge the integral over \textbf{q}. 

Our model makes the following assumptions: (i) Changes induced by the electron-phonon coupling in the electronic wavefunctions and phonon dynamical matrices are small and can be neglected \citep{EPW_package_PONCE2016116}, (ii) similarly renormalization of phonon frequencies due to anharmonic effects is ignored \citep{RevModPhys}, (iii) electron and phonon baths are at the same temperature $T$ all the time. The electron-phonon scattering time for a carrier in the state $|n\mathbf{k}\rangle$ is then given by 
\begin{equation}
  \tau_{n\mathbf{k}}(T)=\frac{\hbar}{2\text{Im}[\Sigma_{n\textbf{k}}(T)]}.\label{eq:tau-nk}
\end{equation}
Conversely, $\text{Im}[\Sigma_{n\textbf{k}}(T)]$ is proportional to the electron-phonon scattering rate $\tau_{n\mathbf{k}}(T)^{-1}$.

\subsection{Time-evolution of excited charge carriers}\label{subsec:method-time-evolution}

We determine the time evolution of the electronic occupation $f_{n\textbf{k}}(t,T)$ using the Boltzmann equation in the RTA
\begin{equation}
  \frac{df_{n\textbf{k}}(t,T)}{dt}=-\frac{f_{n\textbf{k}}(t,T)-f^{\mathrm{th}}_{n\textbf{k}}(T)}{\tau_{n\textbf{k}}(T)}, 
  \label{eq:Boltzmann}
\end{equation}
which has the solution
\begin{equation}
    f_{n\textbf{k}}(t,T)=f^{\mathrm{th}}_{n\textbf{k}}(T)+e^{-\frac{t}{\tau_{n\textbf{k}}(T)}}[f_{n\textbf{k}}(0,T)-f^{\mathrm{th}}_{n\textbf{k}}(T)].
    \label{eq:Boltzmann-solution}
\end{equation}
Here, $f^{\mathrm{th}}_{n\textbf{k}}(T)$ is the thermalized Fermi distribution at $t \to \infty$, where the quasi-Fermi level is chosen to reproduce the number of carriers initially present. The description through the Boltzmann equation~(\ref{eq:Boltzmann}) is valid only, if the number of carriers excited at $t=0$ represents a weak perturbation. RTA relaxes an excited charge carrier directly to the thermalized state, omitting all intermediate relaxation steps between the initial nonequilibrium and final relaxed occupation.

We start with a hot-carrier occupation $f_{n\textbf{k}}(0,T)$ that consists of the sum of a Fermi-Dirac distribution $f^{(0)}_{n\textbf{k}}(T)$ at temperature $T$ and a Gaussian peak centered at energy $+\zeta$ or $-\zeta$ for electrons or holes, respectively, see also Fig.~\ref{fig:heterostructure}, as 
\begin{eqnarray}
     f_{n\textbf{k}}(0,T) & = & f^{(0)}_{n\textbf{k}}(T)-\frac{\lambda_{\text{h}}}{\sqrt{2\pi\sigma^{2}}}e^{\frac{(\varepsilon_{n\textbf{k}}+\zeta)^{2}}{2\sigma^{2}}},\quad \varepsilon_{n\textbf{k}}<\varepsilon_{\text{F}}, \nonumber\\
     f_{n\textbf{k}}(0,T) & = & f^{(0)}_{n\textbf{k}}(T)+\frac{\lambda_{\text{e}}}{\sqrt{2\pi\sigma^{2}}}e^{\frac{(\varepsilon_{n\textbf{k}}-\zeta)^{2}}{2\sigma^{2}}},\quad \varepsilon_{n\textbf{k}}\geq\varepsilon_{\text{F}}.
     \label{eq:initial-distribution}
\end{eqnarray} 
The Gaussian distribution uses a small energy broadening $\sigma = 8.47$~meV. Since valence and conduction bands feature an energy-dependent density of states (DOS), $\lambda_{\text{e}}$ and $\lambda_{\text{h}}$ are adjusted such that the same number of photoexcited carriers is present at any excess energy $\zeta$. 

We define the thermalization time $\tau_{\mathrm{th}}$ as
\begin{equation}
    \frac{P(\zeta,\tau_{\mathrm{th}},T)}{P(\zeta,0,T)}=\frac{1}{e}
    \label{eq:tau_th}
\end{equation}
with the energy-, time- and temperature-dependent population
\begin{equation}
    P(E,t,T)=\sum_{n\mathbf{k}}\delta(E-\varepsilon_{n\mathbf{k}})\times
    \begin{cases}
    [1-f_{n\mathbf{k}}(t,T)],\quad E<\varepsilon_{\text{F}},\\
    f_{n\mathbf{k}}(t,T),\quad E\geq\varepsilon_{\text{F}}.
    \end{cases}
    \label{eq:population}
\end{equation}
In our numerical calculations, we approximate the delta function in Eq.~(\ref{eq:population}) by a narrow Gaussian with a width of 20 meV. We focus on electronic excitations far enough above the quasi-Fermi levels such that the finite width does not affect $\tau_\mathrm{th}$.

\subsection{Interlayer charge transport}

Having the thermalization time at hand, we need a reference timescale to see whether high-energy photoexcited carriers can contribute to interlayer electron transport in vdW heterostructures, as sketched in Fig.~\ref{fig:heterostructure}. An accurate calculation of the interlayer transport time is a challenging task, because the carrier motion across interfaces is subject to multiple uncontrolled effects, {\it e.g.} interfacial roughness and impurity scattering. These effects may vary even within the same batch of samples, let alone the use of different 2D materials. In what follows, we employ a concept based on the transmission coefficient for carriers tunneling through a barrier and the uncertainty relation between the photocarrier excess energy and lifetime \cite{ma2016tuning}. The advantage of the approach is that we can straightforwardly relate the transmission probability of the photoexcited carriers to a measurable quantity, namely the IQE. 

Our starting point is the well-known transmission probability of carriers through a rectangular barrier of width $d$ and height $\Delta_0$ counted from the respective band edge, see Fig.~\ref{fig:heterostructure}. We assume the tunneling regime \cite{ma2016tuning} so that the photocarrier excess energy is always below the barrier $\zeta < \Delta_0$. Note that according to the conventions used in this work $\zeta$ and $\Delta_0$ are both positive, even for holes. We assume that the in-plane photocarrier momentum is not conserved due to interfacal disorder, rendering the problem effectively one-dimensional. We must, however, change the physical meaning of the incident wave vector $k_z$. It does not describe propagating waves anymore but is related to the out-of-plane momentum uncertainty $\Delta p_z/\hbar$, which is of the order of the size of the first Brillouin zone for 2D conductors \cite{APL2018thermionic}. The approximations can formally be summarized as \cite{landau3,ma2016tuning}
\begin{eqnarray}
    \mathcal{T}(\zeta) &=&\frac{4 k_z^2 \kappa^2}{\left(k_z^2 + \kappa^2 \right)^2\sinh^2{\kappa d}
    +4k_z^2 \kappa^2}\\
    &\approx &   \frac{4 k_z^2 \kappa^2}{\left(k_z^2 + \kappa^2 \right)^2}e^{-2\kappa d},\quad
    \kappa d \gg 1 \\
    &\approx &  \frac{4 \kappa^2}{k_z^2}e^{-2\kappa d}, \quad k_z \gg \kappa\\
    & \approx & 8\frac{\Delta_0-\zeta}{v_z \Delta p_z}\exp\left(-\frac{2d}{\hbar}\sqrt{2m(\Delta_0-\zeta)}\right).
\end{eqnarray}
In the last line, we have used the relations $\hbar\kappa=\sqrt{2m(\Delta_0-\zeta)}$ and $\hbar^2 k_z^2/m\approx v_z \Delta p_z$. Since the structure is aperiodic along the out-of-plane direction, the effective mass $m$ of the quasiparticle moving across the interface equals the free electron mass. Importantly, there is a relation between the quasiparticle velocity, momentum uncertainty, and lifetime \cite{landau3,APL2018thermionic}: $v_z\Delta p_z \approx 1/\tau_\mathrm{th}$. Note that we have assumed here that the lifetime is determined by the thermalization time. Hence, the photocarrier transmission probability can be expressed as $\mathcal{T}(\zeta)\approx \tau_\mathrm{th}/\tau_\mathrm{tun}$, where
\begin{equation}
    \frac{1}\tau_\mathrm{tun} 
    \approx \frac{\Delta_0-\zeta}{\hbar}
    \exp\left(-\frac{2d}{\hbar}\sqrt{2m(\Delta_0-\zeta)}\right).
    \label{eq:tunneling}
\end{equation}
    
Equation~(\ref{eq:tunneling}) is somewhat similar to the transport time formula for a triangular barrier, employed in Ref.~\cite{ma2016tuning} to describe electron tunneling in the Fowler-Nordheim regime. In our case, we assume a low bias and a thin barrier such that the voltage does not appear in $\tau_\mathrm{tun}$ explicitly, but it may influence $\Delta_0$. As we need $\tau_\mathrm{tun}$ solely for comparison with $\tau_\mathrm{th}$ by the order of magnitude, we estimate it roughly for $\Delta_0\approx 1$~eV and $d\approx 1$~nm. The interlayer transport time then ranges between 100~fs to 10~ps, depending on the excess energy. To gain a substantial photocurrent, the interlayer transport must not be too slow as compared with thermalization. From the experimental point of view, the ratio $\tau_\mathrm{th}/\tau_\mathrm{tun}$  is nothing else but the IQE in the tunneling limit $\tau_\mathrm{th} \ll \tau_\mathrm{tun}$. In the following section we will further analyze $\tau_\mathrm{th}$, $\tau_\mathrm{tun}$ and the ratio $\tau_\mathrm{th}/\tau_\mathrm{tun}$.

\section{Results}

\begin{figure}[!tb] \centering{}\includegraphics[width=1.0\columnwidth]{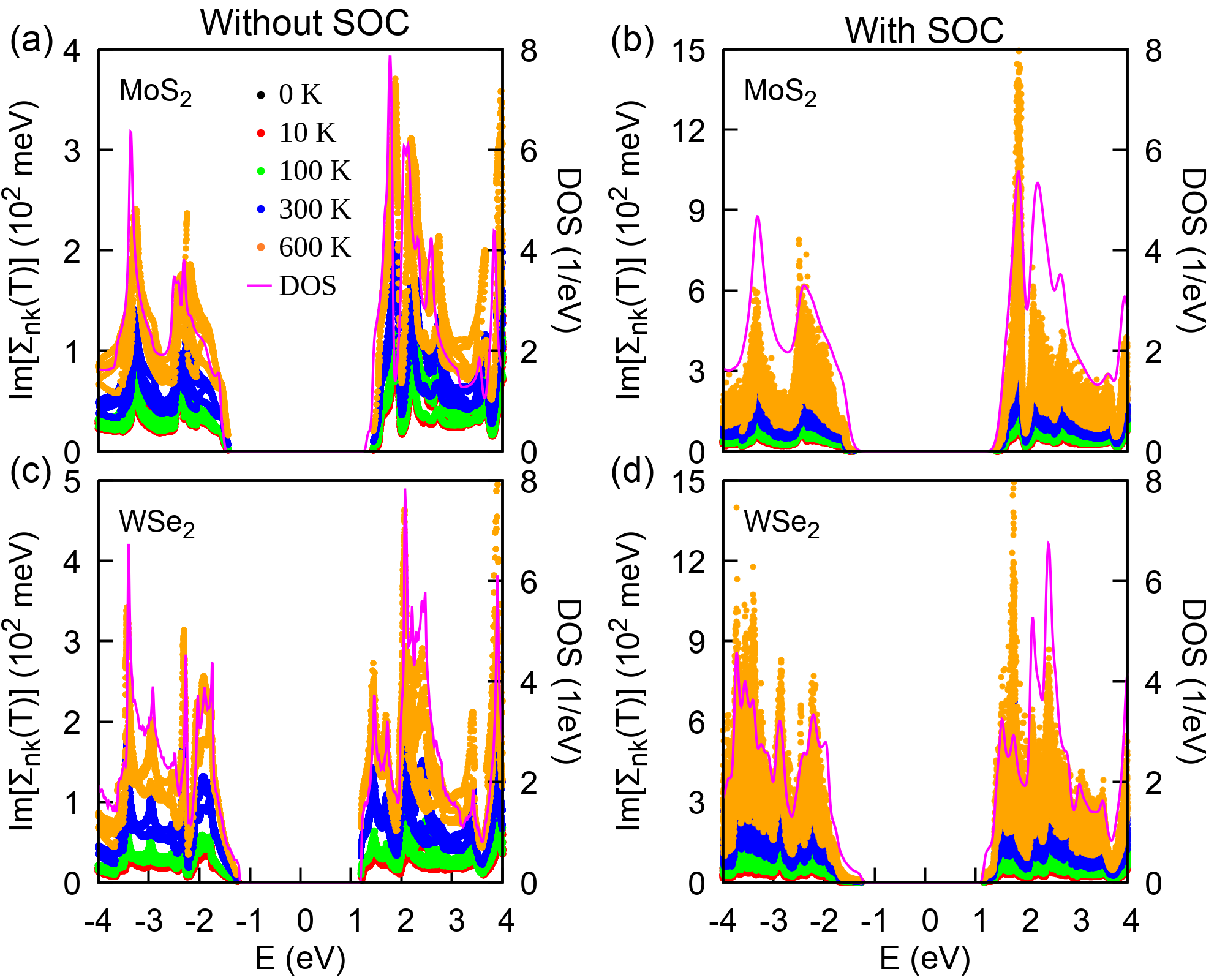}
\caption{Imaginary part of the electron-phonon self-energy as a function of energy calculated for different temperatures without and with SOC for (a)-(b) MoS$_2$ and (c)-(d) WSe$_2$ monolayers. The DOS is also shown in each panel by a solid line.} 
\label{fig:Imsig}
\end{figure}

We start by discussing the imaginary part of the self-energy, which is proportional to the electron-phonon scattering rate $\tau_{n\mathbf{k}}^{-1}$. Figure~\ref{fig:Imsig} shows Im$\selfenergy$ for MoS$_2$ and WSe$_2$ with and without SOC as a function of energy at different temperatures. The energy dependence of Im$\selfenergy$ follows that of the electronic DOS. This can be understood from Eq.~(\ref{eq:self-energy}), where the self-energy for state $|n\textbf{k}\rangle$ is determined by a sum over all electronic states $|m\textbf{k}+\mathbf{q}\rangle$. Considering that phonon energies are limited to below 60~meV for both materials, the sum essentially constitutes an integral over electronic states in the vicinity of $\epsilon_{n\mathbf{k}}$, which is proportional to the electronic DOS. SOC reshapes the electronic band structure by splitting spin-degenerate states. On the rather large energy scales shown in Fig.~\ref{fig:Imsig}, comprising excess energies of up to 3~eV, we find that this can increase the imaginary part of the self-energy as compared to the case without SOC.

\begin{figure}[!tb] \centering{}\includegraphics[width=1.0\columnwidth]{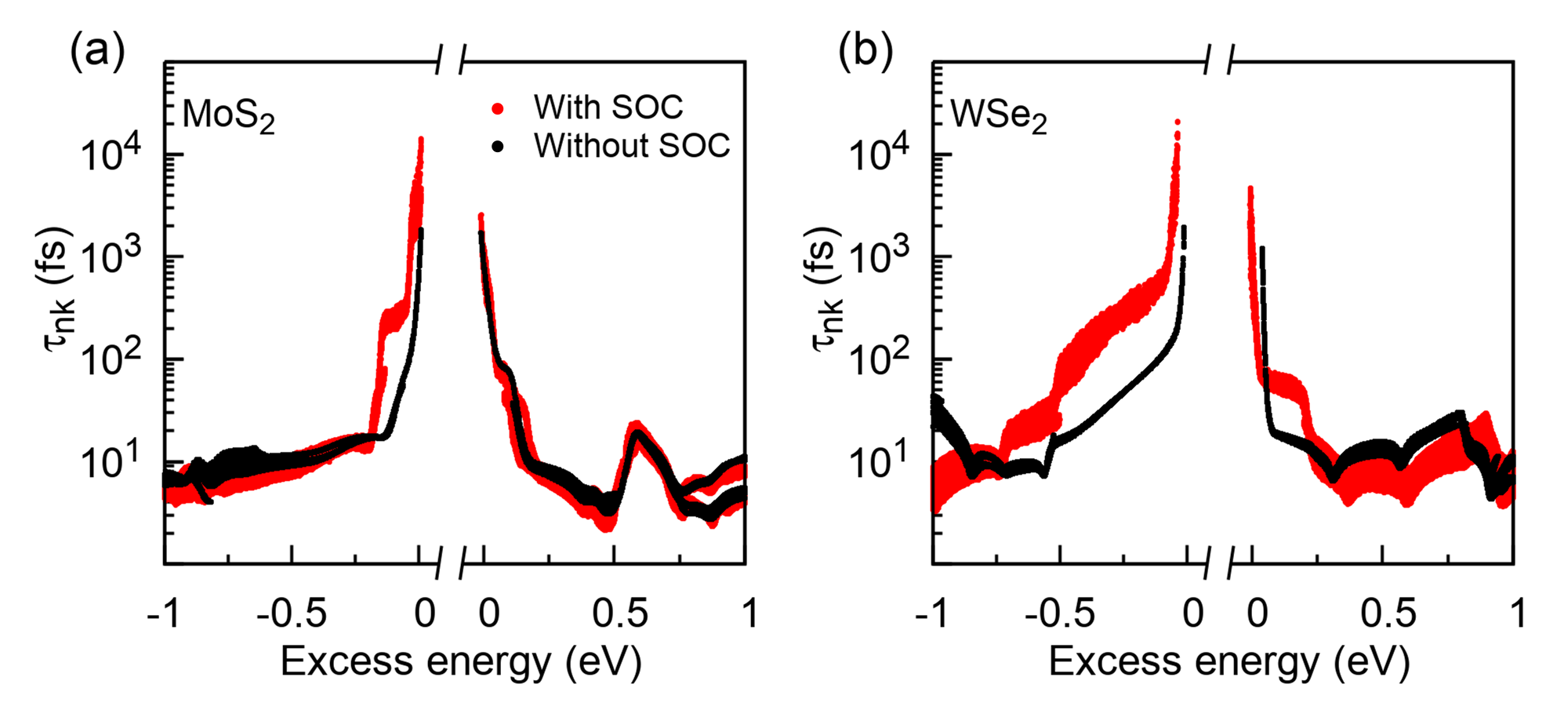}
\caption{Scattering times as a function of excess energy with and without SOC for (a) MoS$_2$ and (b) WSe$_2$ at $T=0$~K.} 
\label{fig:taunk}
\end{figure}

More relevant to us are the effects of SOC around the band edges. Comparing scattering times with and without SOC in Fig.~\ref{fig:taunk}, we observe that the $\tau_{n\mathbf{k}}$ are strongly increased by SOC for holes of both MoS$_2$ and WSe$_2$ and electrons of WSe$_2$. The electron states of MoS$_2$, on the other hand, are basically unaffected. The large increase of $\tau_{n\mathbf{k}}$ correlates with a large spin-orbit splitting of the corresponding valence and conduction band states in Fig.~\ref{fig:bandstruct-e-ph}. Let us point out that similar calculations of $\tau_{n\mathbf{k}}$ have already been presented by Ciccarino \textit{et al.} \cite{Neglect_e_e_interaction_TMDC}. The authors have assigned the increased scattering times to the suppression of intraband scattering and spin-valley locking. In the following we extend that study by calculating thermalization times with the Boltzmann equation, by distinuishing the roles of acoustical and optical phonons in the scattering processes, and by comparing thermalization to tunneling times in vdW heterostructures.

\begin{figure}[!tb] \centering{}\includegraphics[width=1.0\columnwidth]{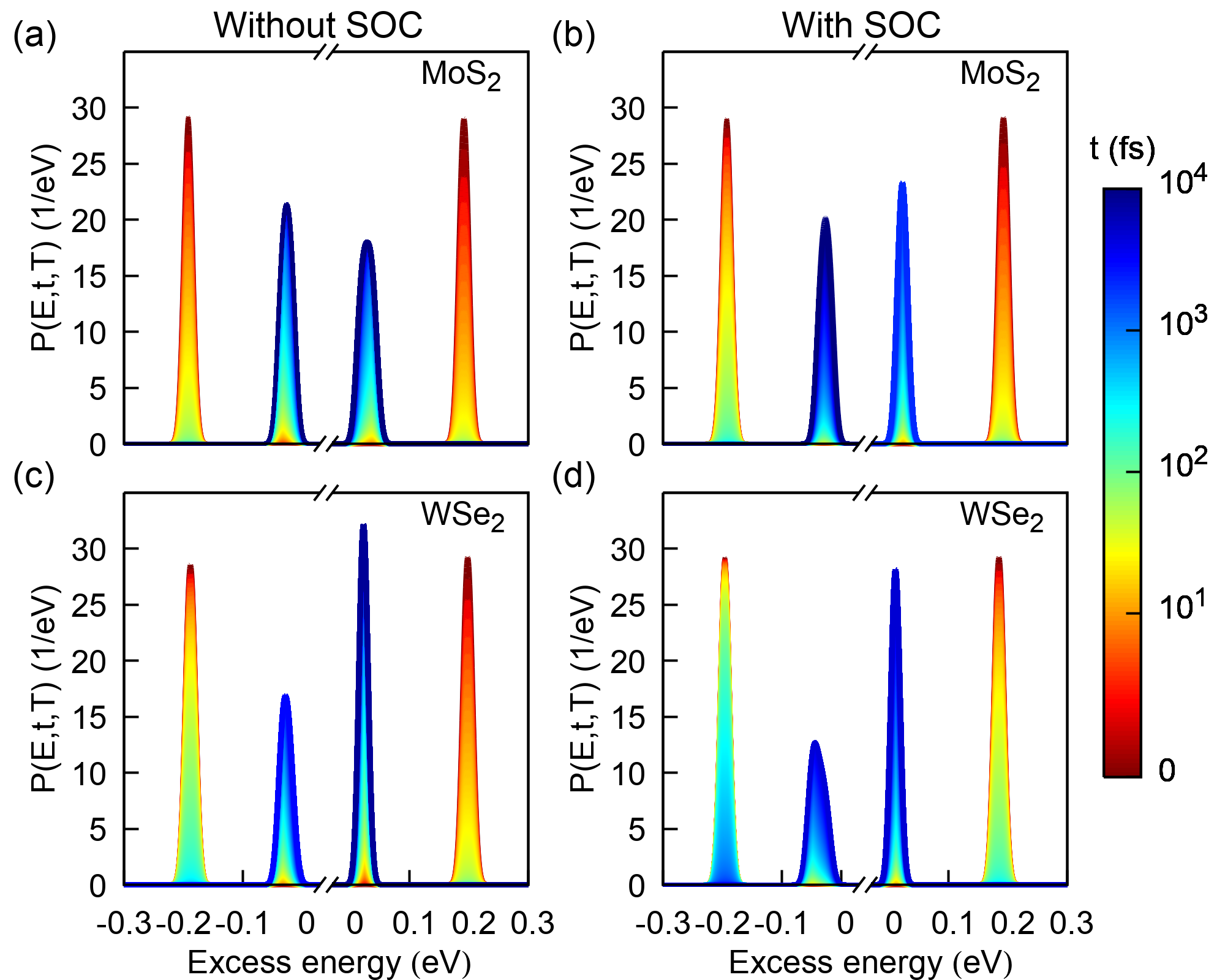}    
\caption{Time evolution of photoexcited carriers with and without SOC in (a)-(b) MoS$_2$ and (c)-(d) WSe$_2$ monolayers at an excitation energy of $\zeta=0.2$~eV for $T=0$~K. Initial and final photoexcited carrier populations correspond to a carrier density of $7\times 10^{12}$~cm$^{-2}$. Note that due to the large spin-orbit splitting near the valence band edge the thermalized population is rather broad in panels (b) and (d), but the area under initial and final distributions remains constant.} 
\label{fig:t-evo}
\end{figure}
    
Typical densities of carriers excited in pump-probe experiments lie between $10^{11}$ to $10^{13}$~cm$^{-2 }$ \cite{Ultrafast_carrier_thermalization_MoS2_1,Exciton_dynamics_wse2,moS2_hetrostructure_extraction}. We adjust our free parameters $\lambda_{\text{e}},\lambda_{\text{h}}$ of Eq.~(\ref{eq:initial-distribution}) such that we generate non-equilibrium populations corresponding to densities of $7\times10^{12}$~cm$^{-2}$ carriers in the valence and conduction bands and use the same number for all the initial state preparations throughout the paper. The time evolution of the occupation is then calculated using Eq.~(\ref{eq:Boltzmann-solution}) with the thermalized Fermi-Dirac distribution $f^{\mathrm{th}}_{n\mathbf{k}}$ centered at a quasi-Fermi level such that the density of $7\times10^{12}$~cm$^{-2}$ carriers is also present after thermalization. This assumption of a conserved number of carriers is valid, because radiative recombination times of electron-hole pairs in MoS$_2$ and WSe$_2$ are on the order of a few ps \cite{Exciton_dynamics_mos2,Bernardi_excitons,Exciton_dynamics_wse2}. They are thus much longer than the thermalization times that we will report in the following.

Fig.~\ref{fig:t-evo}(a) shows the evolution of the hot-carrier population for MoS$_2$ at $T=0$~K without SOC. In this case thermalization times of electrons and holes are comparable. Upon introducing the SOC, it can be seen in Fig.~\ref{fig:t-evo}(b) that the thermalization time of holes is increased, whereas it remains almost unchanged for electrons. WSe$_2$ follows the same trend as MoS$_2$, see Fig.~\ref{fig:t-evo}(c,d). We attribute this behavior to the strong spin-orbit splitting of the valence band in both MoS$_2$ and WSe$_2$.
     
\begin{figure}[!t] \centering{}\includegraphics[width=1.0\columnwidth]{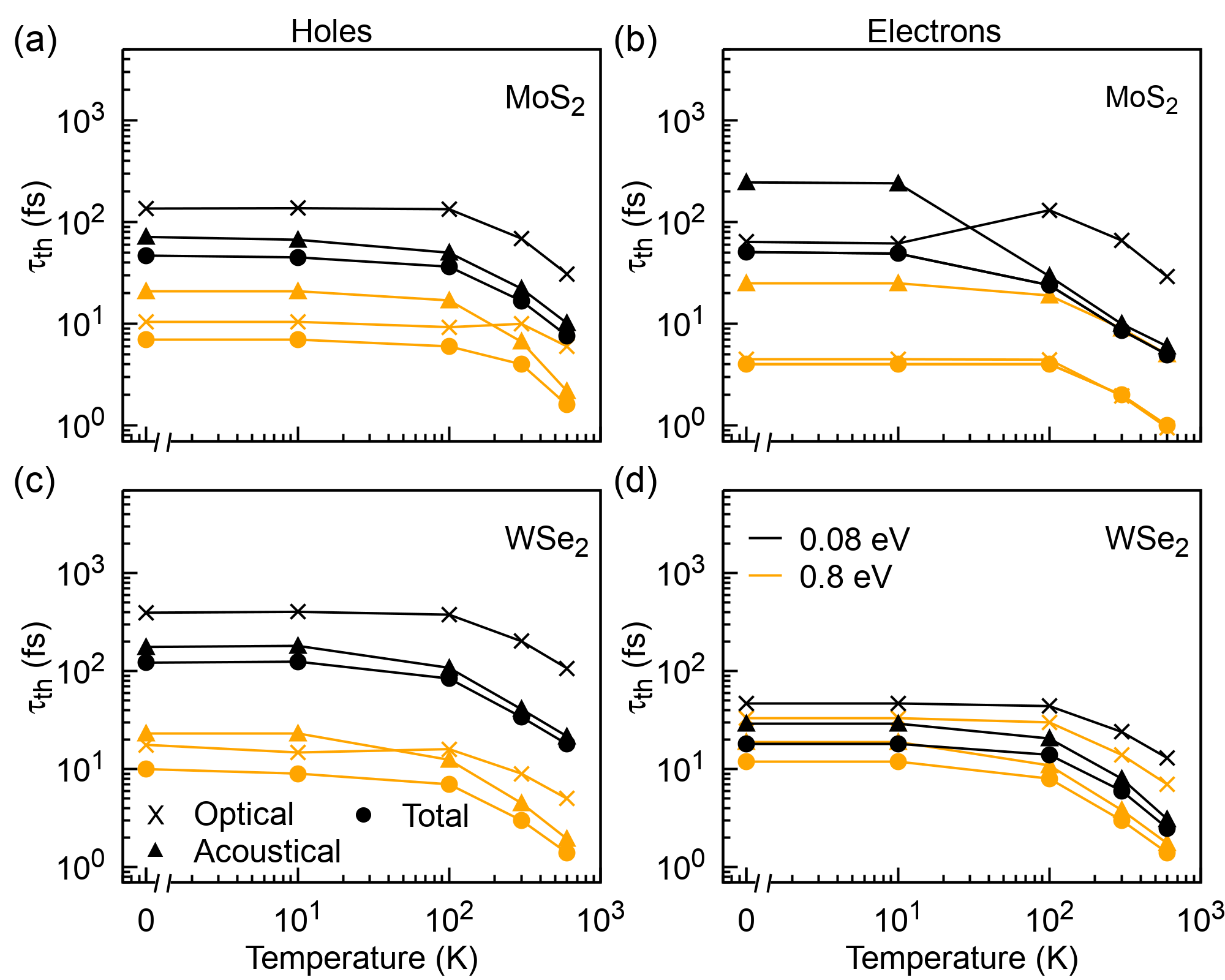}
\caption{Contributions of acoustical and optical phonons to the thermalization of hot holes (left) and hot electrons (right) in (a)-(b) MoS$_2$ and (c)-(d) WSe$_2$ monolayers, neglecting SOC.}  \label{fig:therm-nosoc}
\end{figure}

\begin{figure}[!b] \centering{}\includegraphics[width=1.0\columnwidth]{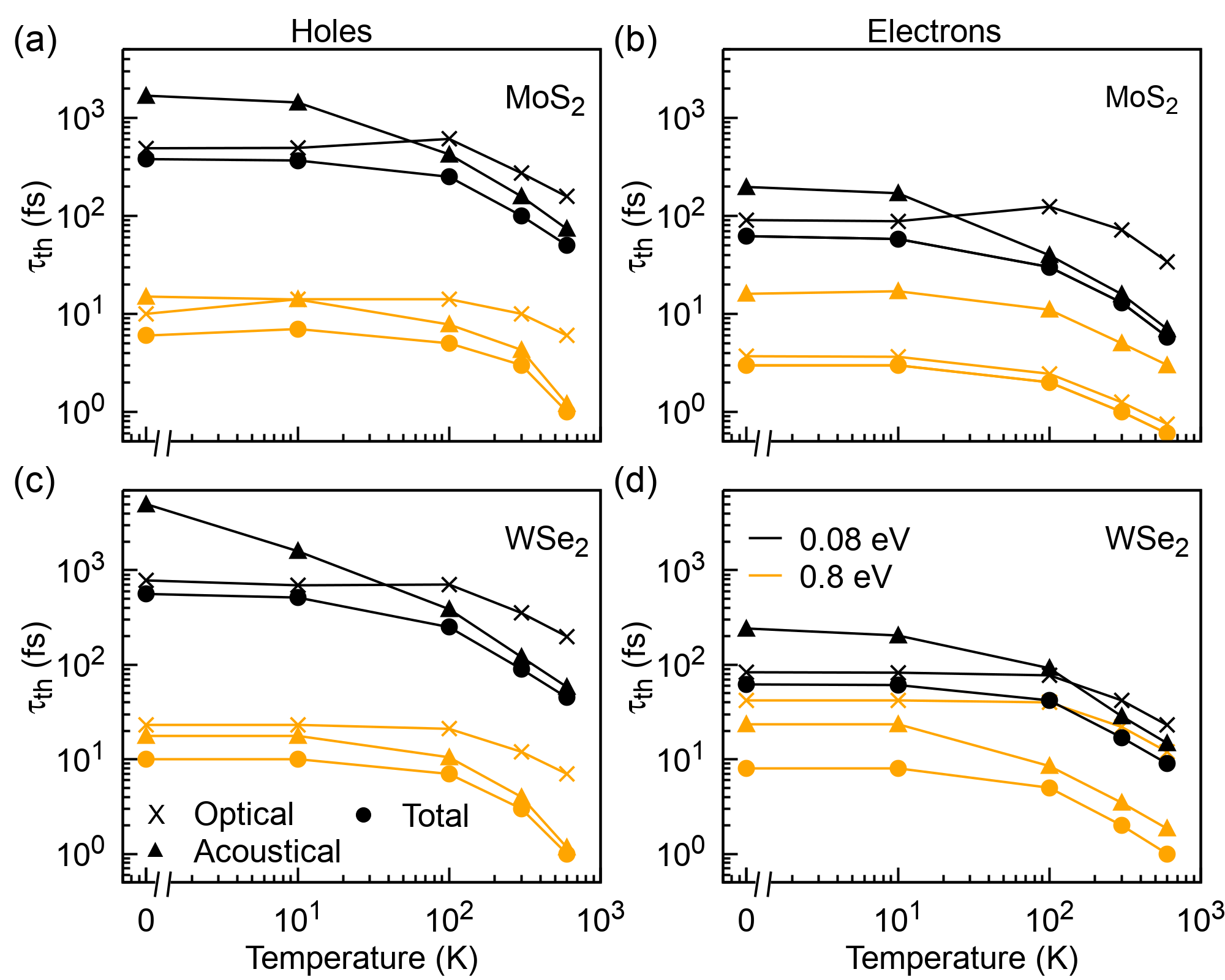}
\caption{Same as Fig.~\ref{fig:therm-nosoc} but including SOC.}
\label{fig:therm-soc}
\end{figure}

The finite energy separation between optical and acoustical branches, observed in Fig.~\ref{fig:bandstruct-e-ph}, makes it possible to distinguish them in $\text{Im}\selfenergy=\text{Im}\selfenergy_{\text{ac}}+\text{Im}\selfenergy_{\text{op}}$ and consequently in $\tau_\mathrm{th}^{-1}=\tau_\mathrm{th,ac}^{-1}+\tau_\mathrm{th,op}^{-1}$. In Figs.~\ref{fig:therm-nosoc} and \ref{fig:therm-soc} we show thermalization times determined by disregarding or considering SOC, respectively. Beside the total thermalization time $\tau_\mathrm{th}$ $(\bullet)$ we display $\tau_\mathrm{th,ac}$ $(\blacktriangle)$ and $\tau_\mathrm{th,op}$ $(\times)$, calculated by taking into account only acoustical or optical phonons. In Fig.~\ref{fig:therm-nosoc} it can be seen that thermalization of electrons and holes at low temperatures near the band edges (see the low excess energy $\zeta = 0.08$~eV) is typically dominated by low-energy acoustical phonons. The electrons of MoS$_2$ constitute an exception, since their thermalization is governed by optical phonons. With increasing temperature acoustical phonons keep their dominant influence on $\tau_\text{th}$ or start to dominante the electron-phonon scattering at $T\gtrsim100$~K for the electrons of MoS$_2$. Away from the band edges (see the high excess energy $\zeta = 0.8$~ eV) a rather diverse picture arises, where both acoustical and optical contributions can define thermalization. Once SOC is introduced in Fig.~\ref{fig:therm-soc}, the contribution of acoustical phonons is strongly suppressed for low $\zeta$ and low $T$, while that of the optical phonons remains nearly unaffected. For this reason the thermalization of both electrons and holes at low temperatures is fully goverened by optical phonons. With increasing temperature acoustical phonons play an increasingly important role and at $T \gtrsim 100$~K they define the decreasing behavior of $\tau_\text{th}$. Altogehter, the effects of SOC slow down thermalization near the band edges.

\begin{figure}[!tb] \centering{}\includegraphics[width=1.0\columnwidth]{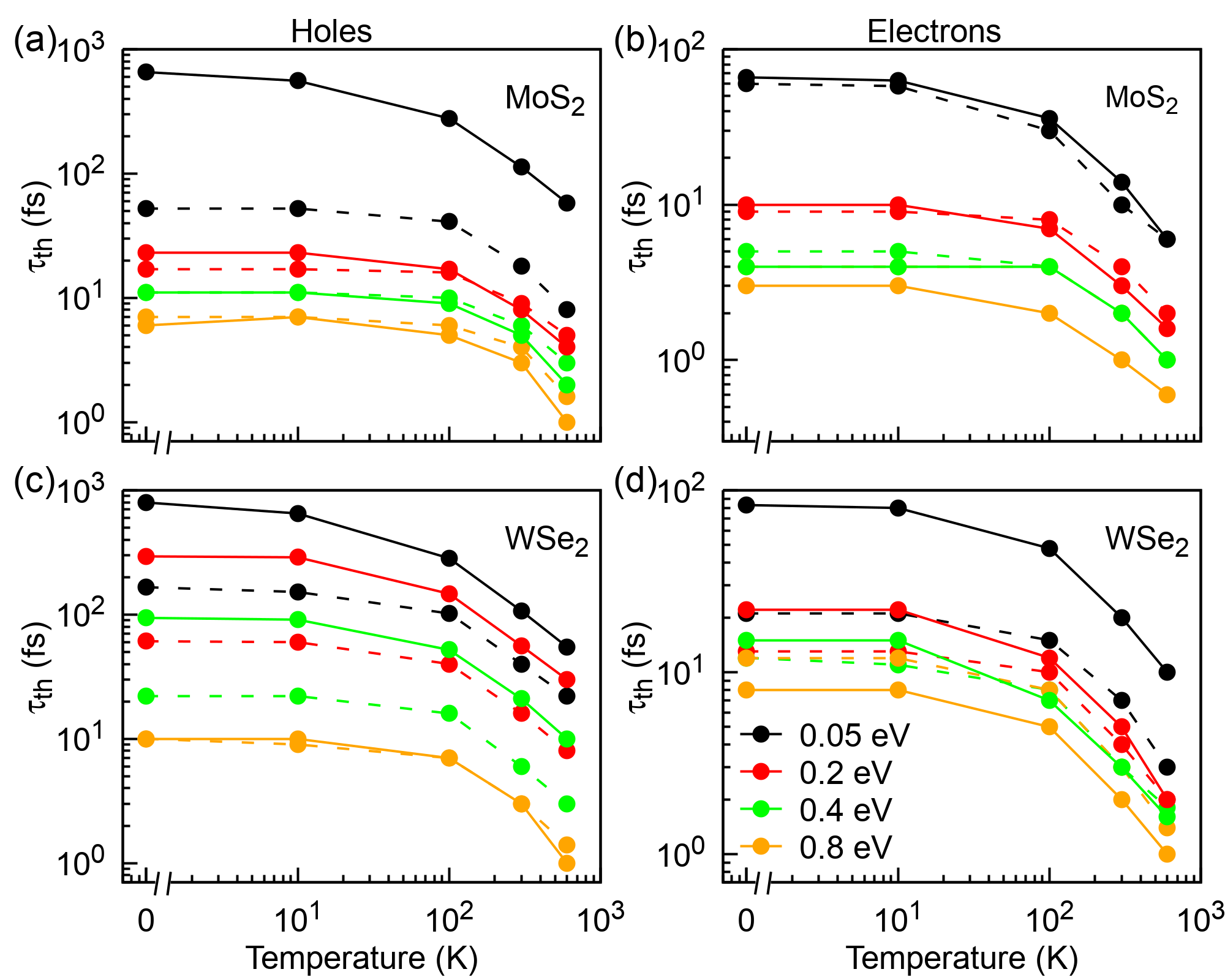}
\caption{Hot carrier thermalization times for holes (left) and electrons (right) in (a)-(b) MoS$_2$ and (c)-(d) WSe$_2$ monolayers. Solid lines represent thermalization times with SOC and dashed lines those without SOC.}  \label{fig:therm-summary}
\end{figure}

We summarize the total thermalization time $\tau_\text{th}$ in Fig.~\ref{fig:therm-summary} as a function of temperature for different excess energies. The dashed curves represent the thermalization time without SOC, and the solid curves include SOC. While SOC strongly increases the thermalization time at low $T$ near the band edges, $\tau_\text{th}$ may also decrease (see for instance $\zeta=0.8$~eV in Fig.~\ref{fig:therm-summary}(d)) depending upon the band structure and electron-phonon couplings. For a fixed excitation energy the thermalization time generally decreases with increasing temperature, because more phonons are accessible to scatter on electrons and holes. The same trend has been reported in pump-probe experiments performed on five-layer MoS$_2$ \cite{Ultrafast_carrier_thermalization_MoS2_1}, where the drop of $\tau_\text{th}$ has been observed for electrons at around $T=300$~K.  Fig.~\ref{fig:therm-summary}(a,b) suggests that it occurs around 100~K for MoS$_2$ in our theory. It should however be emphasized that we study perfectly crystalline, free-standing layers in vacuum. The presence of a substrate, defects and interlayer interactions thus complicates a comparison with the experimental results. A qualitative agreement is nevertheless achieved.
    
As visible in Fig.~\ref{fig:therm-summary}, free carriers near the band edges feature longer relaxation times as compared with those at higher excess energies. The reason is the lower electronic DOS available for scattering around the band edges. Caused by multiple valleys in the conduction band, see Figs.~\ref{fig:bandstruct-e-ph} and \ref{fig:Imsig}, electrons thermalize faster than holes in both MoS$_2$ and WSe$_2$. Due to the heavier elements composing WSe$_2$, the SOC-related splitting at both the valence and conduction band edges is larger than in MoS$_2$. This alters the thermalization time of both electrons and holes. However, electrons still thermalize one order of magnitude faster than holes around band edges. We note that our calculated thermalization times of electrons in MoS$_2$ for $\zeta= 0.05$~eV amount to 63 and 36~fs at 10 and 100~K, respectively, which is of a similar magnitude as found in Ref.~\cite{Carrier_dynamics_free_carriers}, where $\tau_\text{th}\approx30$~fs at $T\approx50$~K.
     
\begin{figure}[!t] \centering{}\includegraphics[width=0.8\columnwidth]{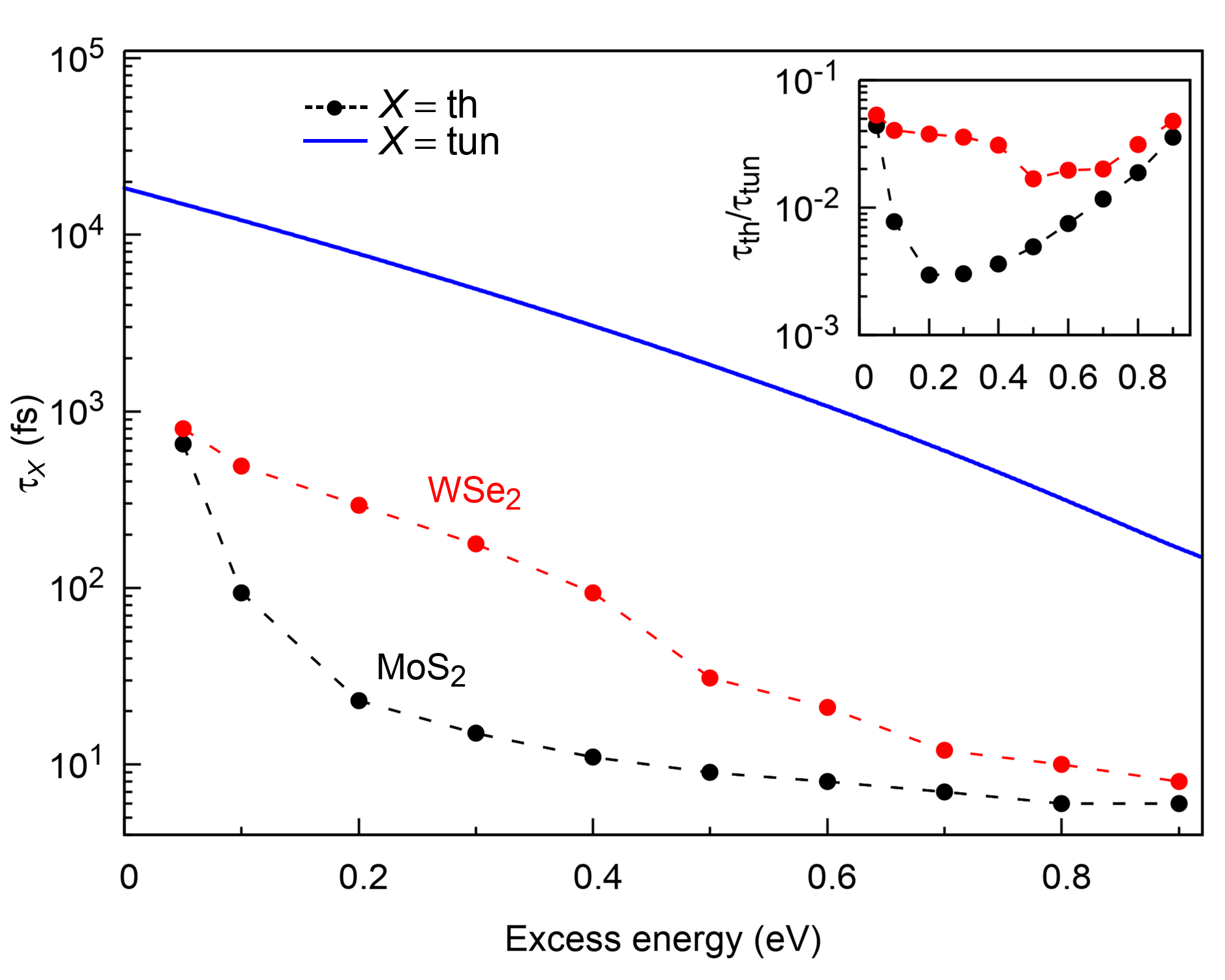}
\caption{Comparison of thermalization and tunneling times for holes in MoS$_2$ and WSe$_2$ heterostructures, see Fig.~\ref{fig:heterostructure}, as a function of excess energy measured from the valence band maxima. The inset shows the ratio  $\tau_{\mathrm{th}}/\tau_{\mathrm{tun}}$. Thermalization times include the effect of SOC.}  \label{fig:therm-tun}
\end{figure}

Since our calculations predict that holes thermalize more slowly than electrons in both MoS$_2$ and WSe$_2$, we focus on hole transport to discuss the IQE of the tunneling device presented in Fig.~\ref{fig:heterostructure}. The heterostructure that we have in mind thus consists of two single layers of 2D TMDCs at the outside, which are separated by a 1~nm thick dielectric film that establishes a tunneling barrier $\Delta_0$ of 1~eV to the valence band maximum of the single layers. Figure~\ref{fig:therm-tun} shows how the photoexcited hole thermalization time $\tau_\mathrm{th}$ compares to the tunneling time $\tau_\mathrm{tun}$ of Eq.~(\ref{eq:tunneling}) as a function of the excess energy $\zeta$. Surprisingly, $\tau_\mathrm{th}$ follows the same trend as $\tau_\mathrm{tun}$ for WSe$_2$. The ratio $\tau_\mathrm{th}/\tau_\mathrm{tun}$ thus depends only weakly on $\zeta$, as visible in the inset of Fig.~\ref{fig:therm-tun}. This suggests that the IQE of the vdW heterostructure stays rather constant with regard to increasing photocarrier excess energy. This contrasts to the case of MoS$_2$, where $\tau_\mathrm{th}$ drops much faster than $\tau_\mathrm{tun}$ when $\zeta$ increases. Hence, we expect that an analogous MoS$_2$-based tunneling device will demonstrate a pronounced decay in the IQE once the photocarriers are excited away from the band edges. Ultimately, our calculations trace back this effect to the larger SOC-induced valence band splitting of WSe$_2$ as compared to MoS$_2$ in these otherwise similar materials, see Fig.~\ref{fig:bandstruct-e-ph}.

\section{Conclusions}
We have combined DFT with many-body perturbation theory to calculate scattering times of photoexcited carriers in MoS$_2$ and WSe$_2$ monolayers arising from electron-phonon interaction. Our model takes all of the phonon branches and their dispersion into account in the scattering processes and highlights the crucial influence of SOC on the thermalization time. In the relevant region near the band edges we find that the inclusion of SOC generally increases the thermalization times by up to an order of magnitude. Our analysis assigns this effect to a suppression of acoustical phonon scattering, while thermalization by optical phonon remains basically unaffected. In both monolayers electrons thermalize almost one order of magnitude faster than holes. We have additionally estimated the tunneling time of a TMDC-based heterostructure using an analytical model. The ratio of thermalization and tunneling times decreases weakly with photocarrier excess energy in WSe$_2$, whereas it drops quickly for MoS$_2$ when moving away from the valence band maximum. Our calculations hence suggest that tunneling devices based on WSe$_2$ monolayers will have a higher internal quantum efficiency than MoS$_2$-based systems.

\acknowledgements

The authors thank G. Eda for stimulating discussions. D.Y.\ and F.P.\ acknowledge financial support from the Carl Zeiss Foundation as well as the Collaborative Research Center (SFB) 767 of the German Research Foundation (DFG).  M.T. acknowledges the Director's Senior Research Fellowship from the Centre for Advanced 2D Materials at the National University of Singapore (Singapore NRF Medium-Sized Centre Programme [R-723-000-001-281], Singapore Ministry of Education AcRF Tier 2 MOE2017-T2-2-140 [R-607-000-352-112], NUS Young Investigator Award [R-607-000-236-133]), and thanks the Okinawa Institute of Science and Technology for its hospitality during two visits of around one week each. Part of the numerical modeling was performed using the computational resources of the bwHPC program, namely the bwUniCluster and the JUSTUS HPC facility.
    
\bibliography{mos2_wse2.bib}
\end{document}